# High Harmonic Spectroscopy Probes Lattice Dynamics

Jicai Zhang[1], Ziwen Wang[1], Frank Lengers[2], Daniel Wigger[2,3], Doris E. Reiter[2,4*], Tilmann Kuhn[2*], Hans Jakob Wörner[5*], & Tran Trung Luu[1*]

[1]Department of Physics, The University of Hong Kong; Pok Fu Lam Rd, Hong Kong SAR, China

[2]Institute of Solid-State Theory, University of Münster; Wilhelm-Klemm-Str. 10, 48149 Münster, Germany

[3]School of Physics, Trinity College Dublin, Dublin 2, Ireland

[4]Condensed Matter Theory, TU Dortmund, 44221 Dortmund, Germany

[5]Laboratorium für Physikalische Chemie, ETH Zürich; 8093 Zürich, Switzerland

*Corresponding author. Email: doris.reiter@tu-dortmund.de; tilmann.kuhn@uni-muenster.de; hwoerner@ethz.ch; ttluu@hku.hk

**Abstract:** The probing of coherent lattice vibrations in solids has been conventionally carried out using time-resolved transient spectroscopy where only the relative oscillation amplitude can be obtained. Using time-resolved X-ray techniques, absolute electron-phonon coupling strength could be extracted. However, the complexity of such an experiment renders it impossible to be carried out in conventional laboratories. Here we demonstrate that the electron-phonon, anharmonic phonon-phonon coupling, and their relaxation dynamics can be probed in real-time using high-harmonic spectroscopy. Our technique is background-free and has extreme sensitivity directly in the energy domain. In combination with the optical deformation potential calculated from density functional perturbation theory and the absolute energy modulation depth, our measurement reveals the maximum displacement of neighboring oxygen atoms in α-quartz crystal to tens of picometers in real space. By employing a straightforward and robust time-windowed Gabor analysis for the phonon-modulated high-harmonic spectrum, we successfully observe channel-resolved four-phonon scattering processes in such highly nonlinear interactions. Our work opens a new realm for accurate measurement of coherent phonons and their scattering dynamics, which allows for potential benchmarking *ab-initio* calculations in solids.

Electron-phonon (*e-ph*, fermion-boson) and phonon-phonon (*ph-ph*, boson-boson) scatterings are two universal interactions in solid matter[1,2]. The former almost entirely determines the optical and electrical properties, and gives rise to new phenomena, e.g., Kohn anomalies[3], the formation of quasiparticles[4], phonon stiffening[5], phonon-assisted absorption[6], etc. It also plays a crucial role in superconductivity[7] and renormalizes electronic excitation energies[8]. The latter originates from the inherent anharmonicity of the chemical bonds in solids and determines important properties of crystals such as lattice thermal conductivity, expansion, infrared, Raman, and neutron scattering cross-sections[9], and closely correlates with interesting effects, such as phonon drag[10], phonon bottleneck[11], and second sound[12]. The advent of ultrafast pump-probe spectroscopy sets the basis to coherently initiate collective atomic motions inside the crystal lattice by photo-absorption of a pump pulse, then detect their electronic and vibrational relaxation processes in real-time through a probe pulse. Over the past few decades, a great number of time-resolved detection techniques have been developed[13]. Generally, coherent phonons can be detected with another ultrashort pulse via transient intensity modulations in reflectivity or transmittivity [14,15]. However, due to the lack of absolute measurement of phonon-induced perturbations, all these methods have limitations in directly quantifying the *e-ph* and anharmonic *ph-ph* interactions.

Since non-perturbative high-harmonic generation (HHG) from solids[16] has been demonstrated following decades of development in strong-field physics and attosecond science from gases[17-19], it quickly attracted much attention from many fields. HHG in solids has been investigated in diverse materials, including bulk insulators[20,21], semiconductors[22], Weyl semimetals[23], metals[24], as well as low dimensional materials[25,26], and the coherent radiation spectrum has been observed from the Terahertz (THz) to the extreme ultraviolet (EUV) range under different driving fields[19-26]. The microscopic mechanism of the HHG process in solids primarily involves two main contributions: intraband current and interband polarization. The latter can be satisfactorily understood by employing a three-step recollision model[27]. The extremely nonlinear nature of the HHG process has opened novel avenues to probe material properties in terms of electronic structure and ultrafast dynamics on sub-femtosecond (1 fs = $10^{-15}$ s) timescales, such as all-optical band structure reconstruction[20], crystal symmetry determination[28], topological character and correlation of materials[29-32], and Berry curvature reconstruction[33]. Furthermore, the high-harmonic spectroscopy (HHS) technique has been utilized to investigate an additional degree of freedom in condensed matter, namely lattice vibrations. While few theoretical studies have proposed the detection of lattice vibrations using

HHS[34,35], the experimental observations of phonon oscillations in VO$_2$ and ZnO systems have been successfully achieved[36,37]. These investigations have predominantly focused on the transient modulation of phonon oscillations within the domain of HHG yields.

Here, we demonstrate for the first time directly in the energy domain, that this extremely sensitive spectroscopy approach can be used to explore *e-ph* and *ph-ph* scattering processes in real-time. Collective periodic lattice vibrations, which can be selectively triggered with a short laser pulse are referred to as coherent phonons[38]. A small displacement $Q$ of the nuclei upon an external driving force $F(t)$, the time-dependent lattice oscillations can be described by the classical equation[39] $\mu\left(\ddot{Q}(t) + \gamma\dot{Q}(t) + \omega_0^2 Q(t)\right) = F(t)$, where $\mu$ is the reduced lattice mass, $\gamma$ the damping constant, and $\omega_0$ the oscillation frequency. In general, $F(t)$ determines the generation mechanism of the coherent phonons that can be classified into a few types, impulsive stimulated Raman scattering (ISRS)[40,41], displacive excitation of coherent phonons (DECP)[42], and resonant excitation mechanisms[43]. As a coherent nuclear displacement $Q(t)$ causes a change in the optical properties (e.g., reflectivity $R$) of the crystal through the refractive index $n$ and the susceptibility $\chi$, the impact of a single mode coherent phonon on the optical response is approximately given by a harmonic response, i.e., $Q(t) \propto \Delta R / R = A_0 \exp(-\gamma t)\cos(\omega_0 t + \varphi)$. This is how a standard transient pump-probe spectroscopy technique detects the dynamics of coherent optical phonons, including four-wave mixing[44], transient reflectivity[14], and transmissivity[15], X-ray diffraction[45,46], as well as second harmonic generation (SHG)[39] spectroscopy methods.

**High-harmonic spectroscopy probes lattice dynamics**

The main idea of initiating and tracking coherent lattice vibrations based on HHG spectroscopy in the dielectric z-cut α-quartz crystal ([0001] direction) is illustrated in Fig. 1. An intense ultrashort pump pulse (~30 fs) and a weaker probe pulse (~25 fs) constitute a non-collinear pump-probe configuration, and time-delayed spectra are recorded by an EUV spectrometer. More details are provided in the Methods part. In principle, when a light pulse interacts with a solid, both direct (resonant) and indirect (non-resonant) excitations of electrons from the valence band (VB) to the conduction band (CB) exist for a small band gap material. The band gap of α-quartz is experimentally measured to be around ~9.5 eV[47] (see calculated band structure in Extended Data Fig. 1) which is much larger than the centre frequency of both pump (~1.55 eV) and probe (~3.1 eV) pulses. Therefore, the creation of electron-hole pairs across the large direct band gap near the Brillouin zone centre only becomes relevant and dominates the transition rate when the laser intensity is sufficiently high to initiate the multiphoton transitions, i.e., when it is close to or higher than $10^{12}$ W/cm$^2$ (nonlinear regime).

As shown in Fig. 1b, the electron transition in a bulk α-quartz crystal happens by an intense laser pulse (~$10^{13}$ W/cm$^2$) primarily through nonlinear multiphoton absorption or electron tunnelling. The coherent driving with two pulses leads to HHG as presented in the Extended Data Fig. 2, both the static HHG spectra of 400 and 800 nm, the harmonics up to 5$^{th}$ of 400 nm and 11$^{th}$ of 800 nm are clearly observed, where the 3$^{rd}$ harmonic has the highest yields. Accompanying electronic transitions, a rapid shift of the crystal potential for the lattice atoms kick-starts coherent phonon oscillations. Conversely, the generated coherent phonon represents a disturbance of the crystal lattice, leading to a variation of the band structure. The photoexcited electron density $n_e(t)$ linearly governs the amplitude of the nuclear shift (coherent phonon amplitude) in the equilibrium coordinate ( $(Q_0'(t)-Q_0)$ ). Note that the timescale of electronic transitions ($10^{-18}$ s) is much faster than typical phonon periods ($10^{-15}$-$10^{-12}$ s), such that the conditions for the (adiabatic) Born-Oppenheimer approximation are satisfied. A time-delayed probe pulse with an intensity of around $10^{12}$ W/cm$^2$ is utilized to detect the coherent phonon dynamics via monitoring the modulation of the time-dependent HHG spectrum of the probe. Figure 1c shows a typical integrated time-delayed 3$^{rd}$ harmonic generation (THG) spectrum of the probe. It obviously exhibits a periodic beating pattern, which is straightforwardly attributed to the coherent lattice vibrations. However, the observed quantity was the time-resolved modulation depth in the spectral domain $\Delta E(t)/E_0$, where $\Delta E(t)=E(t)-E_0$ is the spectral deviation from the HHG spectrum without phonons $E_0$ ('reference spectrum' when comparing to a typical transient spectroscopy measurement). Note that this significantly differs from conventional intensity modulation techniques in transmission ($\Delta T/T$) or reflection ($\Delta R/R$)[14,15,44-46]. It should also be emphasized that our HHG-based technique is background-free.

As illustrated in Extended Data Fig. 3a and 3c, distinct phonon oscillations were also observed in the 4$^{th}$ and 5$^{th}$ harmonics of the probe pulse, and all the modulations show the same phases and frequencies. While no periodic oscillation was observed in the pump pulse when we switched the time delay of the two pulses. In comparison to the THG signal, other harmonics of the probe exhibit a relatively low signal-to-noise ratio (SNR) in the time-resolved spectra. Therefore, unless otherwise specified, our subsequent discussions will solely focus on the THG spectrum of the probe. It should be clarified that by combining the X-ray diffraction and photoemission spectroscopy techniques[46], one can track the *e-ph* interaction strength in real time and real space, yet the complexity of the experimental setup makes it not possible to be carried out in conventional laboratories. In short, our phonon detection method based on HHS offers absolute amplitude information in the energy domain, outperforming current laboratory-based techniques that can only provide amplitude information up to an unknown proportional constant.

To determine the lattice vibration modes and their symmetries, a fast Fourier transform (FFT) of Fig. 1c is performed (Fig. 1d), and two main peaks located at $207.8 \pm 0.7$ cm$^{-1}$ and $464.8 \pm 0.4$ cm$^{-1}$ are found. In comparison to Raman spectra[15] and calculated phonon dispersions based on self-consistent density functional theory (DFT) (see Extended Data Fig. 4), we can assign the two peaks to optical phonons of $A_1$ symmetry, rotation of $SiO_4$ tetrahedra ($A_{1g}$), and a ring $O_{bridge}$-breathing ($A_{1b}$) mode. This analysis further indicates that the electronic state of the α-quartz crystal adiabatically follows the combination of $A_{1g}$ and $A_{1b}$ modes. A direct visualization can be found in the Extended Data Movie. 1. The apparent absence of the ground optical phonon mode $E_{128}$ (located at ~128 cm$^{-1}$) confirms the DECP generation mechanism[42] for the two observed optical phonon modes. In contrast, previous work on α-quartz crystals was mainly focused on the $E_{128}$ mode that is generated by the ISRS mechanism[41]. Once optical phonon states are populated (i.e., "hot" phonons are created), the phonon perturbation not only contributes to the *e-ph* interaction but can also be scattered anharmonically to other phonon branches via three and higher phonon scattering processes as sketched by the waved arrows in Fig. 1b. Within a time-frequency analysis using the Gabor transform (GT, see Methods part) of the observed spectra (Fig. 1c), as shown in Fig. 1e ($A_{1g}$) and Fig. 1f ($A_{1b}$), clear frequency modulation spectra of the two hot phonon modes are discovered, which grants us to additionally trace the anharmonic *ph-ph* scattering dynamics on a fs timescale. Note that the GT sacrifices spectral resolution but allows us to keep temporal information in a spectral representation.

**Quantum-classical model**

The α-quartz crystal has three high symmetry points, Γ, K, and M in the Brillouin zone (see Extended Data Fig. 4b). Through HHG measurements in solids one can acquire the symmetry information of the crystal[28], such that we can select the laser polarization parallel to the high symmetry path Γ-M or Γ-K via rotating the orientation of the crystal. Figure 2a shows the integrated THG trace and its centre of mass (COM, solid black line) of probe that is detected when the laser is polarized in the Γ-K direction. To extract the spectral dynamics of the observed two hot phonon modes ($A_{1g}$ and $A_{1b}$), we perform a GT of the spectrum that yields instantaneous frequencies and amplitudes of the phonon oscillations. Figure 2b shows the GT corresponding to the COM of the detected spectrum (Fig. 2a), evidently showing the strong modulations in the COM line at the frequency centres of the $A_{1g}$ and the $A_{1b}$ mode. The modulation depth (peak-to-peak) of the $A_{1g}$ and the $A_{1b}$ mode is found to be ~60 cm$^{-1}$ and ~14 cm$^{-1}$, respectively.

To develop a deeper understanding of the time evolution of the physical processes of the THG radiation trace of the probe pulse, following the findings that the optical properties in the region close to the band gap are strongly dominated by excitonic effects[47], we numerically solve the following two-level quantum model describing the optically driven and phonon-coupled exciton (detail see Methods, Dynamical Quantum Model for HHG Radiation Involving Phonons)

$$\frac{d}{dt} f_0 = -2\varepsilon(t)\,\text{Im}[p] \tag{1}$$

$$\frac{d}{dt} p = -i\Omega(t)p + i\varepsilon(t)(1-2f_0) - \frac{p}{T_2}, \tag{2}$$

with the exciton occupation $f_0$, the polarization $p$, the instantaneous Rabi frequency of the driving electric field $\varepsilon(t)$, the time-dependent exciton energy $\Omega(t)$ and the dephasing time $T_2$. We want to remark that the terms $\varepsilon(t)\,\text{Im}(p)$ and $\varepsilon(t)f_0$ in eq. (1-2) give rise to a nonlinear, mutual coupling of the two equations which is the reason why such optical Bloch equation-based models can be used to describe HHG processes[48,49]. We examine the Fourier transform of the microscopic polarization as the source of the emitted radiation $E(\omega) \propto p(\omega)$. Due to the *e-ph* interaction, the excitonic variables are coupled to the phonon dynamics. While incoherent phonons give rise to exciton-phonon scattering processes and phonon-induced dephasing, coherent phonons lead to a modulation of the exciton energy according to the time-dependent coherent phonon amplitude[50], as will be shown in more detail in the Methods section. To quantitively determine the strength of *e-ph* and *ph-ph* interactions, we combine the two-level quantum model with a classical coupled oscillator interaction model by considering that the phonon coupling leads to a time-dependent exciton energy as[51,52]

$$\hbar\Omega(t) = E_{1S} + V_{1b}\cos\left[\omega_{1b}(t)t + \varphi_{1b}\right]\exp\left[-\frac{t}{\tau_{1b}}\right] + V_{1g}\cos\left[\omega_{1g}(t)t + \varphi_{1g}\right]\exp\left[-\frac{t}{\tau_{1g}}\right] \quad (3)$$

$$\omega_{1g}(t) = \omega_{1g}\left[1 + \gamma_1\cos(2\omega_1 t + \varphi_1) + \gamma_2\cos(\omega_2 t + \varphi_2)\right] \quad (4)$$

$$\omega_{1b}(t) = \omega_{1b}\left[1 + \gamma_1'\cos(2\omega_1' t + \varphi_1') + \gamma_2'\cos(\omega_2' t + \varphi_2')\right], \quad (5)$$

Where $E_{1S}$ represents the unperturbed exciton energy, $V_i$ the *e-ph* interaction strength, $\omega_i$ and $\omega_i'$ the centre phonon oscillation frequencies, $\tau_i$ the phonon decay time, $\gamma_i$ and $\gamma_i'$ the dimensionless anharmonic *ph-ph* coupling constants, $\varphi_i$ and $\varphi_i'$ the initial phonon phases. Note that in order to quantify the anharmonic *ph-ph* scattering strength by using the classical coupled oscillator model[51], an acoustic ($\omega_1$ and $\omega_1'$) and an optical phonon ($\omega_2$ and $\omega_2'$) are considered (see following *ph-ph* scattering). The result of the simulated spectral dynamics is shown in Fig. 2c and its GT in Fig. 2d. With properly fitted parameters of phonon modes (for the details, see the Least-Squares Fitting in the Methods part), we achieve remarkably good agreement with the experimental results. The time-resolved even, and higher odd harmonic spectra traces also can be accurately reproduced, as demonstrated in Extended Data Fig. 3b and 3d, by taking into account a complex transition dipole moment[33].

**Optical manipulation of *e-ph* coupling dynamics**

To quantitatively measure the impact of the $A_{1g}$ and $A_{1b}$ hot phonon scattering dynamics by the light pulses, we fix the electric field strength of the probe pulse to ~0.15 V/Å and record the time delayed THG spectra under different pump intensities (~0.76 to 1.14 V/Å). Figures 3a-3c show the spectral dynamics for pump intensity scaling, where the periodic oscillation amplitude linearly rises with increasing pump fluence. We also find that the modulation amplitude does not grow symmetrically, while the minima remain almost unaffected, the maxima shift to higher energies. This implies that the lattice vibration-induced perturbation has an anisotropic character when interacting with VB and CB. Indeed, as shown in Fig. 3d, the exciton energy determined with our model, reflecting the density-dependent band gap, shows a linear dependence on the pump fluence with a slope of 21.8 ± 4 meV/(V/Å). This is very surprising because the band gap usually shows a well-known decreasing trend with increasing temperature in thermodynamics theory[53]. This unusual discrepancy indicates that, in a highly nonlinear system where both 'hot' electrons and phonons are involved in light-matter interactions, the contribution of the *e-ph* coupling dominates the process for the pump influence rather than thermal expansion[54] on such short time scales, and inversely for the probe influence (see Extended Data Fig. 5).

In addition to the constant shift of the band gap, as displayed in Fig. 3e, the *e-ph* interaction strengths of the $A_{1g}$ and $A_{1b}$ modes show a linear growth with slopes of 52.7 ± 8 and 31.7 ± 7 meV/(V/Å). This is reasonable because the coherent phonon amplitude increases with a higher pump flux, which then results in a larger modulation amplitude of electron and hole energies. To figure out the main reasons, we recorded for the probe intensity scaling as shown in Extended Data Fig. 6, where for the whole range scaling is more appropriately fitted with linear rather than the perturbative power law fitting, implying the light-matter interaction is close to the non-perturbative regime (at least at higher intensity). One usually expects that the phonon relaxation time (or lifetime) monotonically drops with increasing temperature of the crystal. Nevertheless, we observed an anomalous phenomenon, as shown in Fig. 3g, the pump dependent relaxation time grows with a slope of 873 ± 82 and 409 ± 73 fs/(V/Å) for the $A_{1g}$ and $A_{1b}$ modes, respectively. One possible explanation is that the *e-ph* interaction changes more actively at higher pump fluences and marginally delays the direct decay channel of the two hot phonon modes. Another significant factor of coherent phonons is the initial phase, which can be interpreted as excitation time to form a specific lattice vibration upon photo absorption. Figure 3h shows the phase variation of the two phonon modes with the pump fluence, indicating that the $A_{1g}$ mode kickstarts the oscillation earlier than $A_{1b}$, and displays a slightly shorter excitation time with higher pump intensity. Based on the measured initial phase and oscillation frequency of the two phonon modes, one can define a relative excitation time $\delta t = \varphi_{1g} / \omega_{1g} - \varphi_{1b} / \omega_{1b}$ (Fig. 3k) between $A_{1g}$ and $A_{1b}$ mode, representing their relative initialization time when absorbing the pump photons in the beginning. The relative delay time between $A_{1b}$ and $A_{1g}$ can be up to 60 ± 10 fs and shows a decrease with higher pump strength with a rate of -25 ± 5 fs/(V/Å).

Moreover, the *e-ph* interaction can be quantified by the optical deformation potential (ODP, see Methods part, Optical Deformation Potential Calculation) from self-consistent density functional perturbation theory (DFPT). For the $A_{1g}$ and the $A_{1b}$ mode the calculated ODP values at the Brillouin zone centre are -0.54 ± 0.15 and 2.31± 0.27 meV/pm, respectively. Applying these values, we retrieve a maximum displacement of neighbouring oxygen (O) atoms in α-quartz crystal to ~27 pm and ~10 pm in real space, respectively (see the Fig. 3f). Associating with the measured *e-ph* strength and calculated ODP, we can access the pulse intensity dependent O atoms displacement of the $A_{1g}$ and $A_{1b}$ mode as shown in Fig. 4a and 4b. In addition, considering the $A_{1g}$ and $A_{1b}$ mode are two orthogonal phonon modes (see Extended Data Fig. 4a and 4b), the 2D oxygen atom displacement trajectories in real space can be conveniently constructed in Cartesian coordinates as shown in Fig. 4c (pump intensity 1.1 V/Å). The vibrational direction of neighbouring base atoms (Si/O) associated with the $A_{1g}$ and $A_{1b}$ modes show almost a π flip in the beginning, which agrees with the calculated vibrational direction and strength from first principles (see Extended Data Fig. 4).

### *ph-ph* scattering process

In the next step, we investigate the anharmonic *ph-ph* coupling dynamics that are encoded in the *e-ph* coupling process. The Extended Data Fig. 7a and 7b display the spectral dynamics of the phonon modes $A_{1g}$ and $A_{1b}$ under different probe peak intensities (0.161 to 0.170 V/Å) and fixed pump (~1 V/Å) with polarization in the Γ-M direction. The modulated amplitudes of the phonon frequencies present an insensitive variation under different probe fluences compared to the *e-ph* interaction in the pump fluence dependence. The extracted parameters from our model are shown in Extended Data Figs. 7c to 7f, where all physical quantities show a very flat tendency when changing the probe fluence, also considering the statistical error bar range. Therefore, here we only quantitatively discuss the averaged interactions for the *ph-ph* coupling dynamics.

The fitted modulation frequencies, as shown in Extended Data Fig. 7c for the $A_{1b}$ and $A_{1g}$ modes, are located at $\omega_1 = 36 \pm 6$ and $\omega_2 = 129 \pm 8$, as well as $\omega_1' = 112 \pm 9$ and $\omega_2' = 258 \pm 8$ cm$^{-1}$, respectively, which agree well with the results from direct FFT (see Extended Data Fig. 8c and 8d). The finding that $\omega_1$ and $\omega_1'$ are even smaller than the ground optical phonon mode $E_{128}$ suggests that these two modes belong to acoustic (ac) phonons. The oscillation frequencies $\omega_2$ and $\omega_2'$ are similar to those of the degenerate optical phonon modes $E_{128}$ and $E_{262}$ of the α-quartz crystal, which indicate the scattered optical phonon modes for $A_{1g}$ and $A_{1b}$ can be $E_{128}$ and $E_{262}$. The obtained anharmonic constants (Extended Data Fig. 7d for the $A_{1g}$ phonon mode), i.e., the scattering strength of $A_{1g}$ with ac and $E_{128}$ modes, are $\gamma_1 = 0.06 \pm 0.02$ and $\gamma_2 = 0.12 \pm 0.03$, respectively. While for the $A_{1b}$ mode, the anharmonic scattering strength with ac and $E_{262}$ modes are $\gamma_1' = 0.007 \pm 0.002$ and $\gamma_2' = 0.01 \pm 0.003$, respectively. Based on the anharmonic coupling analysis of the two $A_1$ phonons, we can conclude that the $A_{1b}$ is more robust and less perturbed by other phonons than $A_{1g}$. The initial phases of the acoustic and optical phonon $E_{128}$ frequencies (Extended Data Fig. 7e) are $\varphi_1 = 0.15 \pm 0.06$ and $\varphi_1' = 4.4 \pm 0.5$ rad for $A_{1g}$, and $\varphi_2 = 1.1 \pm 0.2$ and $\varphi_2' = 1.1 \pm 0.3$ rad for $A_{1b}$, which results in corresponding relative delay times of $\delta t = 22 \pm 9$ fs for the $A_{1g} \to ac$ and $A_{1g} \to E_{128}$, and $\delta t' = 68 \pm 23$ fs for the $A_{1b} \to ac$ and $A_{1b} \to E_{262}$ scattering processes (Extended Data Fig. 7f). From these measurements we can conclude: (1) Even though a coherent phonon mode is excited early and decays faster, the coherent *ph-ph* scattering process is still present and can be traced by a faster attosecond scale electronic motion; (2) usually in a quantum picture the *ph-ph* scattering description is instantaneous, but the time for the build-up of a collective lattice oscillation varies among the different vibrational modes from a classical perspective.

In accordance with the extracted modulation frequencies, together with a reliable phonon dispersion relation (see Extended Data Fig. 4c and Tab. 1), we can assign the previous results to possible *ph-ph* scattering pathways using the following four-phonon scattering selection rules[1]

$$q_1 = q_2 + q_3 + q_4 + G \quad \text{(Type I)} \tag{6}$$

$$q_1 + q_2 + q_3 = q_4 + G, \quad \text{(Type II)} \tag{7}$$

where the four phonons participating in the scattering process have wave vectors $q_1, q_2, q_3$ and $q_4$ and $G$ is a reciprocal lattice vector to account for normal ($G = 0$) and umklapp ($G \neq 0$) processes. The α-quartz has three acoustic and twenty-four optical phonon branches. Due to the limited time window of the phonon oscillation signal, it is difficult to directly resolve the lower frequency of the modulated phonon mode. From the obvious spectral modulation peak of the fundamental phonon mode, as shown in Extended Data Fig. 8a, a regular spacing frequency around 130 cm$^{-1}$ is observed. There is no splitting in the spectrum indicating that the degenerate mode $E_{128}$ at q = 0 participates in the scattering process. The almost continuous spectrum between 0 and 100 cm$^{-1}$ implies that the other scattered phonons are likely continuous mode phonons, i.e., acoustic phonons. Similarly, we can assign the $E_{262}$ and acoustic modes as the scattering process based on the peak spacing in the GT spectra of the $A_{1b}$ mode as presented in Extended Data Fig. 8b. Based on the regular spacing frequency analysis and the selection rules of the conservation of phonon energy and quasi-momentum, the most likely phonon scattering is a process (*1*) and belongs to 1st type scattering, thus the assignment of the possible pathways for $A_{1g}$ and $A_{1b}$ phonons are indicated by the arrows in Fig. 5.

Since we analyse the *ph-ph* scattering directly in the time-domain, the time dependent scattering rate, as illustrated in Extended Data Fig. 8a and 8b, can be monitored from a time-frequency analysis of the spectral dynamics in Extended Data Figs. 7a and 7b, respectively. The scattering processes reach maxima at times around 430 fs ($E_{128}$) and 350 fs (ac) for $A_{1g}$, and 0 ($E_{262}$) and 150 fs (ac) for $A_{1b}$. Such a time-frequency analysis of the *ph-ph* interaction allows us to extract information on the connected phonon frequencies in real time. Taking the short-lived $A_{1g}$ mode as an example, as shown in Extended Data Fig. 8a, the center frequency of the excited ac phonon shifts towards lower energies with time, indicating that the $A_{1g}$ phonon either scatters to a lower ac phonon branch or that the ac phonon decays along its phonon dispersion curve. Thus, comparing to the conventional angle-resolved photoemission spectroscopy technique[46], an alternative all-optical way to measure the phonon dispersion curve could be developed from here, i.e., from a hot *ph-ph* scattering process and the considered simple time-frequency analysis. Note that there have been some slight deviations between the observed phonons and DFT calculations, to access a full phonon dispersion relation with higher accuracy to benchmark calculations from the first principles, for instance under cryocooling conditions, a high SNR spectrum and long time-window of phonon oscillations should be measured and analysed.

**Conclusion**

All-optical triggering and probing lattice vibration dynamics using HHS directly in the energy domain establishes a new paradigm for direct measurements of *e-ph* and *ph-ph* couplings. Our technique not only provides benchmark data for theoretically determining fundamental physical quantities describing the couplings from first principle's calculation (e.g., DFT). The linearly dependent intensity scaling law of *e-ph* coupling dynamics also reveals the feasibility to optically control the electronic, phononic, and their interaction properties in condensed matter with millielectronvolt (meV, lattice potential) and picometre (pm, real space) precisions. The identification and analysis of channel-resolved four-phonon scattering processes using a simple and reliable time-windowed Gabor analysis method present a valuable opportunity to further explore the intricate and intriguing anharmonic phonon-phonon coupling processes in complex systems. Beyond α-quartz, 2D crystals such as graphene[55] and transition-metal dichalcogenides (TMDs)[56] as a new generation of electronic functional device materials, will benefit from accurate measurements of the fundamental *e-ph* and *ph-ph* interaction dynamics, which will shed some insights on their carrier mobility, heat transfer, and consequently applications in the future.

**Methods**

**Material**

In the experiment, we utilize a z-cut α-quartz crystal (100 orientation) (United Crystal) of 5*5 mm size as target for high-order harmonics generation (HHG). Its surface was two-sided optically polished, and the thickness was measured to be around 20 μm with our homemade white light interferometry spectrometer. The crystal structure of α-quartz is made of a continuous framework of Si–O tetrahedron ($SiO_4^{4-}$), and each oxygen atom is shared by two tetrahedrons. The unit cell consists of 9 atoms in total as shown in Extended Data Fig. 1a. According to group theory, α-quartz has a trigonal crystal system and belongs to the space group $P3_221$ (right-handed)[57]. The resulting phonon modes are divided into three acoustic vibrations ($A_2 + E$) and twenty-four optical vibrations of $4A_1 + 4A_2 + 8E$ symmetry. The non-degenerate $A_1$ modes are Raman active, the $A_2$ modes are infrared active, and the doubly degenerate E modes are both infrared and Raman active.

**Experimental Setup and HHG Spectroscopy Technique**

The experimental setup is shown in Fig. 1a of the main text, two linearly polarized (P) laser pulses constitute the non-collinear time-delayed pump-probe HHG spectral detection geometry. The non-collinear angle is less than 1.5 degree. The pump stems from a high-power Ti: Sapphire near-infrared (NIR) laser at the carrier wavelength of 800 nm with a total energy of 7 mJ, and a repetition rate of 1 kHz. The probe pulse (400 nm) is obtained by frequency-doubling the 800 nm pulse. The pulse durations of the pump and probe are determined by our homemade transient grating frequency-resolved optical gating (XEng Limited) setup to around 30 fs and 25 fs, respectively. A 75 cm focal length lens is used to focus the two light pulses on the sample. Note that due to the different opening sizes of the iris, the resulting focal beam size on the sample of pump and probe is around 150 - 200 μm, and 60-80 μm, respectively. The size of the probe beam at the focus is roughly half that of the pump beam due to the lower power input (and also the Rayleigh criterion for diffraction limits). After strong interaction with the sample, the HHG signal will be generated along the same direction of the two-driving pulses. Then through a slit for filtering the fundamental pulses, the high order harmonics signal is separated by a flat-field variable groove density grating (Hamamatsu), and each harmonic is amplified and recorded by the CCD camera (PCO Panda) coupled micro-channel plates (MCP) detector. The spectral range reaches from 5 eV to 55 eV, limited by the collection angle of the EUV spectrometer. Note that the resolution of the EUV spectrometer is around 0.05 eV but can be improved further via the higher statistical averages. The spectrum is not corrected for the sensitivity of the grating. When performing the time delay scan, a linearly closed-loop piezo stage is used for precisely detuning one of the arms of the pump and probe. Also, using Fresnel's formula for S and P polarization, under a normal incident condition together with power-camera measurements, the intensity amplitude of the pump and probe inside the sample are estimated to be around $10^{13}$ and $10^{12}$ W/cm$^2$, respectively. Note that our EUV spectrometer has a limited capability to resolve phonon modulation depths, with a maximum resolution of up to 0.001 eV. Additionally, it can only measure phonon oscillations within a restricted time window of less than 1.5 ps. These limitations are due to the rapid decay time of phonon modes and the limited range of our piezo stage.

**Fast Fourier Transform and Gabor Transform**

When performing the time-resolved spectral analysis, we make use of the Fast Fourier Transform (FFT) and the Gabor Transform (GT). The FFT, converts a signal from its original domain *F(t)* (usually time or space) to a representation in the reciprocal domain $\phi(v)$ (frequency or momentum, respectively) and vice versa. It is given by the following well-known formula as[58]

$$F(t) = \int_{-\infty}^{\infty} \phi(v) e^{2\pi i v t} dv \qquad (8)$$

$$\phi(v) = \int_{-\infty}^{\infty} F(t) e^{-2\pi i v t} dt. \qquad (9)$$

The GT is a special case of the short-time FFT and it is usually utilized to determine the sinusoidal frequency and phase content of local parts of a signal that changes over time. The function to be transformed is first multiplied by a (modified) Gaussian function, which can be regarded as a window function *g(t)*, and the resulting function is then transformed with a Fourier transform to derive the time-frequency analysis when varying the center of the window function[59]. A peaked window function leads to a higher weight of the signal at the time being analysed. The Gabor transform is defined as

$$S(\omega, \tau) = \int_{-\infty}^{\infty} s(t) g(t-\tau) \exp(-i\omega t) dt, \qquad (10)$$

where $s(t)$ is the time-dependent signal and $g(t-\tau)$ a super-Gaussian window function reading

$$g(t-\tau) = \exp(-(t-\tau)^n / \tau_p^n). \qquad (11)$$

To achieve a high-resolution time-dependent spectrum, it is crucial to optimize the time width and the order *n* of the super-Gaussian window function. In this study, two phonon modes, namely the $A_{1g}$ and $A_{1b}$ modes, were observed at approximate phonon wave numbers of 207 cm$^{-1}$ and 464 cm$^{-1}$, respectively. The corresponding phonon oscillation periods for the $A_{1g}$ and $A_{1b}$ modes are 161 fs and 72 fs, respectively. In order to monitor the anharmonic phonon-phonon coupling between these two modes in real-time, we employed a time-window of 160 fs, *n* = 6 for the time-delayed HHG spectrum, and 350 fs, *n* = 8 for the phonon frequency modulation spectrum, respectively. To clearly explain the influence of the window width influence, as one can see in the Extended Data Fig. 9, with the decreasing of the window width from 150 to 50 fs, there will be no dynamic feature of $A_{1g}$ mode on the spectra.

**Least-Squares Fitting**

All the fittings we used in the present are based on the Curve Fitting Toolbox in MATLAB software, which uses least-squares fitting methods to estimate the coefficients of a regression model. The algorithm for calculating the vector of estimated responses is

$$\hat{y} = f(X,b).$$
(12)

Where the $\hat{y}$ response estimates, $f$ is the general form of the regression model. $X$ is a design matrix. $b$ is the parameters of fitted model coefficients. A least-squares fitting method is employed to compute model coefficients that minimize the sum of squared errors (SSE), also known as the residual sum of squares. For a given set of n data points, the residual for the $i^{th}$ data point is calculated using the following formula:

$$SSE = \sum_{i=1}^{n}(y_i - \hat{y}_i)^2.$$
(13)

The fitting result as in Extended Data Fig. 10, which excellently agrees with the experimental data, with a residual error of less than 1%. To assess the robustness of our fitting, we conducted an analysis where we intentionally set the pre-values of the $A_{1b}$ and $A_{1g}$ phonon modes to 200 and 100 cm$^{-1}$, respectively, with lifetimes of 800 and 200 fs. These values resulted in all four parameters being off by 50%. Even with such significant deviations, as shown in Extended Fig. 10 b) as long as we employed a larger number of iterations (~15,000) based on our fitting algorithm, the fitting still achieved a very low residual error. The excellent agreement between the experimental observations and the fitted results suggests that the main contribution to the modulation in the central energy of the harmonics arises from phonon modulations. In contrast, the propagation effect, or phase matching, appears to be weak and negligible.

**Density Functional Theory**

Ab initio density functional theory (DFT) is employed to determine the electron band structure and density of states (DOS) of the α-quartz crystal. The band structure and DOS shown in Extended Data Figs. 1c and 1d were calculated on a commercial platform: Quantum Atomistix ToolKit (ATK) Q-2019.12[60] based on first principles methods. In the calculation, geometry optimization was done under the force field approximation, with the force tolerance reached the level of $10^{-3}$ eV/Å. and the following electronic structure calculations were done within the meta generalized gradient approximation (MGGA) in the parametrization of Perdew-Burke-Ernzerhof (PBE). In the calculation, the TB09 functional and a high-accuracy Pseudo-Dojo basis set were used, and the energy cut-off was set to 830 eV, the iteration tolerance is $10^{-6}$. A Γ–centered Monkhorst-Pack of 6 x 6 x 5 was applied in the Brillouin zone. By using the TB09 functionals and the Pseudo-Dojo pseudopotential, we determine the direct band gap (i.e., the band gap at the Γ point) to 9.3 eV, which agrees well with the experimentally measured value of 9.5 eV[61-63].

**Density Functional Perturbation Theory**

Phonon quantities, such as phonon dispersion curves, vibrational modes, and optical deformation potentials (related to the electron-phonon coupling) are calculated by density functional perturbation theory (DFPT) using the frozen phonon method of lattice displacements[2] on the same ATK platform[59]. As all these calculations are correlated to the dynamical matrix calculation, the lattice constants were selected as $a$ = 4.9160 Å and $c$ = 5.4054 Å upon a zero-pressure optimized structure using the Limited-memory Broyden–Fletcher–Goldfarb–Shanno (LBFGS) optimizer method. The calculations were performed with the GGA-PBE variant of the GGA exchange-correlation functional. The pseudopotentials were chosen as the optimized norm-conserving variety that was generated by the GGA method and gave a converged calculation result at a cut-off energy of 1250 eV. A 11×11×9 Monkhorst-Pack mesh was used to calculate the electronic Brillouin-zone integrals, which converged the computation at a force constant less than 1 meV/Å. A 7×7×5 supercell (corresponding to 2205 atoms in total) and a 5 pm atomic displacement of Si and O atoms in the 3D real space were used for the dynamical matrix and Hamiltonian derivatives calculations, and the iteration tolerance is $10^{-10}$. Especially when performing the calculation of phonon dispersions as shown in Extended Data Fig. 4, we directly used the previously published classical force field constants[64] which gave the best comparable results with the experimental Raman spectroscopy measurements [65,66] until now.

**Dynamical Quantum Model for HHG Radiation Involving Phonons**

The band gap of α-quartz lies around 9.5 eV[61-63] and it also exhibits strong excitonic features around the band gap[47], which dominate the optical response in this frequency range. The central frequency of the probe pulse (3.1 eV) is far away from a resonant band gap transition and high above the THz-region where polarization and current sources of the HHG are of the same order. Therefore, we conclude that the optical signals are dominated by excitonic transitions rather than by quasiparticle band-to-band transitions. Since the oscillator strength of higher excitonic transitions is quickly decaying with increasing exciton quantum number $n$ ($\sim n^{-3}$ in the case of bulk Wannier excitons[67]), we will restrict ourselves to a single exciton level with $n = 1$, i.e., the 1s-exciton, which is most strongly coupled to the light field. Introducing the creation and annihilation operators $X_{\mathbf{K}}$ and $X_{\mathbf{K}}^{\dagger}$ for 1s-excitons with center-of-mass wave vector $\mathbf{K}$ [49,68], the free exciton Hamiltonian then reads

$$H_X = \sum_k E_{\mathbf{K}} X_{\mathbf{K}}^{\dagger} X_{\mathbf{K}} \tag{14}$$

with the corresponding energies $E_{\mathbf{K}} = E_{1S} + \hbar^2 K^2 / 2M$ with the total exciton mass $M$. Due to the far off-resonant excitation, high-density effects can be neglected, and the exciton operators therefore fulfil bosonic commutation relations $\left[ X_{\mathbf{K}}, X_{\mathbf{K}'}^{\dagger} \right] = \delta_{\mathbf{KK}'}$. We consider the excitonic transition driven by an electric field $E(t) = E_0 \cos(\omega t)$ coupled via the dipole moment matrix element along the direction of the electric field $d_0$, which we assume to be real. Note, that we do not perform the rotating wave approximation. Assuming a homogeneous system with a homogeneous excitation, only excitons with $K = 0$ couple to the electric field resulting in the exciton-light coupling Hamiltonian

$$\begin{aligned} H_{X-light} &= -d_0 E(t) X_0^{\dagger} - d_0 E(t) X_0 \\ &= -\hbar \varepsilon(t)(X_0^{\dagger} + X_0), \end{aligned} \tag{15}$$

where we have introduced the abbreviation $\varepsilon(t) = d_0 E(t) / \hbar$ denoting the instantaneous Rabi frequency of the driving field. To describe the coupling of the excitons to the phonons we introduce the bosonic phonon creation and annihilation operators $B_{i,\mathbf{Q}}^{\dagger}$ and $B_{i,\mathbf{Q}}$ for a phonon with wave vector $\mathbf{Q}$ and energy $\hbar \omega_{i,\mathbf{Q}}$ in branch $i$, as well as the exciton-phonon coupling matrix element $g_{\mathbf{Q}}^{(i)}$. Then the free phonon Hamiltonian and the Hamiltonian of the exciton-phonon interaction read

$$H_{ph} + H_{X-ph} = \sum_{i,\mathbf{Q}} \hbar \omega_{i,\mathbf{Q}} B_{i,\mathbf{Q}}^{\dagger} B_{i,\mathbf{Q}} + \sum_{i,\mathbf{K},\mathbf{Q}} \hbar g_{\mathbf{Q}}^{(i)} X_{\mathbf{K}+\mathbf{Q}}^{\dagger} X_{\mathbf{K}} (B_{i,-\mathbf{Q}}^{\dagger} + B_{i,\mathbf{Q}}). \tag{16}$$

With this Hamiltonian we can set up the equations of motion for any expectation value via the Heisenberg equation of motion

$$\frac{d}{dt}\langle A \rangle = \frac{i}{\hbar}\langle [H, A] \rangle, \tag{17}$$

which for the exciton occupation $f_K = \langle X_K^{\dagger} X_K \rangle$ leads to the equation

$$\frac{d}{dt}f_{\mathbf{K}} = -i\varepsilon(t)(p-p^*)\delta_{\mathbf{K},0} +$$
$$i\sum_{\mathbf{Q},i}g_{\mathbf{Q}}^{(i)}\left(\langle X_{\mathbf{K+Q}}^{\dagger}X_{\mathbf{K}}B_{i,-\mathbf{Q}}^{\dagger}\rangle + \langle X_{\mathbf{K+Q}}^{\dagger}X_{\mathbf{Q}}B_{i,\mathbf{Q}}\rangle - \langle X_{\mathbf{K}}^{\dagger}X_{\mathbf{K-Q}}B_{i,-\mathbf{Q}}^{\dagger}\rangle - \langle X_{\mathbf{K}}^{\dagger}X_{\mathbf{K-Q}}B_{i,\mathbf{Q}}\rangle\right), \quad (18)$$

for the polarization $p = \langle X_0 \rangle$ to

$$\frac{d}{dt}p = -i\frac{E_{1s}}{\hbar}p + i\varepsilon(t)(1-2f_0) + i\sum_{\mathbf{Q},i}g_{\mathbf{Q}}^{(i)}\langle X_{-\mathbf{Q}}B_{i,-\mathbf{Q}}^{\dagger}\rangle + \langle X_{-\mathbf{Q}}B_{i,\mathbf{Q}}\rangle, \quad (19)$$

and for the coherent phonon amplitude $\langle B_{i,\mathbf{Q}}\rangle$ to

$$\frac{d}{dt}\langle B_{i,\mathbf{Q}}\rangle = -i\omega_{i,\mathbf{Q}}\langle B_{i,\mathbf{Q}}\rangle - i\sum_{\mathbf{K}}g_{\mathbf{Q}}^{(i)}\langle X_{\mathbf{K+Q}}^{\dagger}X_{\mathbf{K}}\rangle. \quad (20)$$

Because of the many-body nature of the coupled exciton-phonon system, we find that higher order terms $\sim \langle X_{-\mathbf{Q}}B_{i,-\mathbf{Q}}^{\dagger}\rangle$ and $\sim \langle X_{-\mathbf{Q}}B_{i,\mathbf{Q}}\rangle$, giving rise to phonon-assisted transitions as well as dephasing due to exciton-phonon scattering, appear in the equation of motion of the polarization $p$. Furthermore, terms such as $\langle X_{\mathbf{K+Q}}^{\dagger}X_{\mathbf{K}}B_{i,-\mathbf{Q}}^{\dagger}\rangle$ describing redistributions in the exciton occupation due to exciton-phonon scattering processes appear in the equation for $f_{\mathbf{K}}$. In order to deal with these terms, we use a factorization scheme according to

$$\langle X_{-\mathbf{Q}}B_{i,-\mathbf{Q}}^{\dagger}\rangle = \langle X_{-\mathbf{Q}}\rangle\langle B_{i,-\mathbf{Q}}^{\dagger}\rangle + \delta\langle X_{-\mathbf{Q}}B_{i,-\mathbf{Q}}^{\dagger}\rangle, \quad (21)$$

$$\langle X_{\mathbf{K+Q}}^{\dagger}X_{\mathbf{K}}B_{i,-\mathbf{Q}}^{\dagger}\rangle = \langle X_{\mathbf{K+Q}}^{\dagger}X_{\mathbf{K}}\rangle\langle B_{i,-\mathbf{Q}}^{\dagger}\rangle + \delta\langle X_{\mathbf{K+Q}}^{\dagger}X_{\mathbf{K}}B_{i,-\mathbf{Q}}^{\dagger}\rangle, \quad (22)$$

which separates the influence of the coherent phonons from exciton-phonon correlations. On the sub-picosecond time scales considered here phonon scattering-induced redistributions are of minor importance, therefore in the lowest order, we neglect the correlations $\delta\langle X_{-\mathbf{Q}}B_{i,-\mathbf{Q}}^{\dagger}\rangle$ and $\delta\langle X_{\mathbf{K+Q}}^{\dagger}X_{\mathbf{K}}B_{i,-\mathbf{Q}}^{\dagger}\rangle$. Since, due to the homogeneous excitation, only excitons with $\mathbf{Q}=0$ are optically excited and the exciton occupation remains diagonal, also coherent phonons are only created with wave vector $\mathbf{Q}=0$ and the phonon-related terms in the equation of motion for the occupation cancel. The equations of motion then simplify to

$$\frac{d}{dt}f_0 = -i\varepsilon(t)(p-p^*), \quad (23)$$

$$\frac{d}{dt}p = -i\left(\frac{E_{1s}}{\hbar} - g_0^{(i)}\left(\left\langle B_{i,0}^\dagger\right\rangle + \left\langle B_{i,0}\right\rangle\right)\right)\left\langle X_0\right\rangle + i\varepsilon(t)(1-2f_0) \quad (24)$$
$$= -i\Omega(t)p + i\varepsilon(t)(1-2f_0),$$

$$\frac{d}{dt}\left\langle B_{i,0}\right\rangle = -i\omega_{i,0}\left\langle B_{i,0}\right\rangle - ig_0^{(i)}f_0. \quad (25)$$

From these equations, it is clear that the exciton-phonon coupling on the one hand leads to the creation of coherent phonons with vanishing wave vectors and on the other hand leads to a dynamical renormalization of the exciton energy via the function

$$\Omega(t) = \frac{E_{1s}}{\hbar} - \sum_i g_0^{(i)}\left(\left\langle B_{i,0}^\dagger\right\rangle + \left\langle B_{i,0}\right\rangle\right). \quad (26)$$

Note, that since only phonons with vanishing wave vector can be excited in a homogeneously excited homogeneous system, only optical phonons are relevant. The renormalization is directly proportional to the displacement of the respective phonon mode given by

$$\left\langle u_i\right\rangle = \sqrt{\frac{\hbar}{2\mu_i\omega_{i,0}}}\left(\left\langle B_{i,0}^\dagger\right\rangle + \left\langle B_{i,0}\right\rangle\right), \quad (27)$$

where $\mu_i$ denotes the reduced mass of the corresponding optical phonon mode $i$. Let us briefly comment on the phonon-induced dynamical energy shift. The typical exciton-phonon coupling as in Eq. (16) is linear in the displacement, the coupling Hamiltonian therefore describes an energy proportional to the lattice displacement operator. In first order perturbation theory this leads to an energy shift proportional to the expectation value of the displacement (see Eq. (26)). In the case of incoherent phonons, e.g., in the presence of thermal phonons, this expectation value is zero and one has to go to second order perturbation theory, where the exciton-phonon correlations in Eq. (21) lead to an energy shift depending on the mean square displacement or, for thermal phonons, on the temperature. In the present case, however, due to the ultrafast pump pulse, coherent phonons are excited, characterized by a non-vanishing expectation value of the lattice displacement. Therefore, we observe in the HHG spectra a non-vanishing energy correction to the exciton energy in first order perturbation theory proportional to the lattice displacement, according to Eqs. (26,27).

In the case of the excitation by optical pulses in the few femtosecond range the driving term in the equation for the coherent phonon amplitudes, resulting from the exciton occupation, acts essentially as a step function, which initiates oscillations around a displaced equilibrium position for each coupled optical phonon mode. This results in the time-dependent energy renormalization in Eq. (26) reading

$$\Omega(t) = \frac{E_{1s}}{\hbar} + \sum_i \frac{V_i}{\hbar} \cos(\omega_{i,0} t + \varphi_i) \exp\left[-\frac{t}{\tau_i}\right], \tag{28}$$

with coefficients $V_i$ being proportional to the amplitude $\langle B_{i,0} \rangle$ of the coherent phonon mode $i$ and we have added a phenomenological damping time $\tau_i$ of the respective phonon mode.

To obtain the optical signals, we numerically solve the coupled set of equations of motion driven by the probe pulse (Eq. (23-25)) (complemented by a phenomenological dephasing time $T_2$ in Eq. (24)) and subsequently calculate the Fourier transform of the microscopic polarization $p$, which acts as the source of the emitted radiation, i.e., $E(\omega) \propto p(\omega)$. In other words, the main microscopic contribution of the high-harmonic generation in this study is the interband polarization, which is similar with the "three-step model" observed when an atom (or molecule) is subjected to a strong light field. Note that the $p(\omega)$ To take into account the effect of anharmonic couplings among the phonons, we replace the frequencies $\omega_{i,0}$ in the energy renormalizations in Eq. (28) according to the coupled oscillator model, as outlined in the main next. For the exciton-phonon coupling constants optical deformation potentials for the valence and the conduction band are obtained from DFT and DFPT calculations, as will be described below.

**Optical Deformation Potential Calculation**

In order to theoretically quantify the *e-ph* interaction strength from first principles, the key task is to calculate the *e-ph* matrix element $g^i_{mn}(\mathbf{K}, \mathbf{Q})$ representing the electronic response following an electron transition process where a Bloch electron from a state with band index $n$ and wave vector $\mathbf{K}$ transitions to a state with band index $m$ and wave vector $\mathbf{K}+\mathbf{Q}$[69-71]. As mentioned previously in the quantum model, here the electronic excitation mainly happens at K = 0, it is reasonable to calculate the optical deformation potential of conduction band and valence band at K = 0 and take their difference as the optical deformation potential of the observed exciton.

The *e-ph* matrix can be determined from the variational formulation in DFPT as

$$g^i_{mn}(\mathbf{K}, \mathbf{Q}) = \sqrt{\frac{\hbar}{2u_i \omega_{i\mathbf{Q}}}} M^i_{mn}(\mathbf{K}, \mathbf{Q}) \tag{29}$$

where $\omega_{i\mathbf{Q}}$ is the specific phonon frequency, and $M^i_{mn}(\mathbf{K}, \mathbf{Q})$ is defined as

$$M^i_{mn}(\mathbf{K}, \mathbf{Q}) = \int_{\mathbf{r}} \psi^*_{m, \mathbf{K}+\mathbf{Q}}(\mathbf{r}) \delta_{i\mathbf{Q}} V(\mathbf{r}) \psi_{n, \mathbf{K}}(\mathbf{r}) d\mathbf{r}, \tag{30}$$

where the initial and final electronic wave functions are extracted from DFT calculations and the perturbation potential $\delta_{i\mathbf{Q}} V(\mathbf{r})$ can be computed by DFPT.

Generally, the electron band structure is determined by the crystal potential and can be influenced by the lattice displacement. For optical phonons, the unit cell of the crystal has two or more atoms, and the neighbouring atoms are displaced in opposite directions (see Extended Data Figs. 4a and 4b). In this case, it is the varying distances between the basis atoms, which disturbs the surrounding lattice potential, serving as a source for the interaction with the electrons. Therefore, the perturbation potential is directly proportional to the oxygen atomic displacement as[70]

$$V_{e-ph} = D_{ODP} u, \qquad (31)$$

where $D_{ODP}$ (eV/pm) is the optical deformation potential (ODP) and $u$ is the atomic displacement. Compared to $V_{e-ph} = M^i_{mn}(0,0) \cdot u_{i\mathbf{Q}}(\mathbf{r},t)$, we can find that the $M^i_{mn}(0,0)$ is directly the $D_{ODP}$. Consequently, $D_{ODP}$ is described as the zero-order deformation potential, which can be calculated as

$$D_{ODP} = M^i_{mn}(0,0). \qquad (32)$$

According to the measured phonon-perturbed THG spectrum, we can directly quantify the *e-ph* interaction strength (perturbation potential). Therefore, after obtaining the deformation potential $D_{ODP}$ value at the Γ point from DFPT of first principles, we can quantify the corresponding atomic displacements for the two phonons in the α-quartz crystal. To compute $D_{ODP}$, we consider the highest valence and lowest conduction band as an initial and final electronic state that can be extracted from DFT calculation. The optical phonon branches are selected as those of the experimentally observed phonon modes. In addition, as the laser polarization is always parallel with a high-symmetry direction (either Γ-K or Γ-M direction) in momentum space, we can directly calculate $D_{ODP}$ along this direction. The $D_{ODP}$ value in the main text is given by the difference of the ODP values of VB to CB at the Γ point. Different XC functions were applied on the calculation and provided the averaged value of the $D_{ODP}$.

**Acknowledgments:** It is our pleasure to acknowledge fruitful discussions with Magnus Molitor. We gratefully acknowledge funding support in part by the ETH Zurich Postdoctoral Fellowship Program (FEL–31 15–2), the Marie Curie Actions for People COFUND Program, and SNSF R'equip grant 206021_170775; in part by the Department of Physics, Faculty of Science, HKU, the RGC ECS project 27300820, the GRF project 17315722, and the Area of Excellence project AoE/P-701/20. D. W. is funded by the Science Foundation Ireland (SFI) under Grant 18/RP/6236.

**Author contributions**: Jicai Zhang & Ziwen Wang constructed the experimental set-up, Jicai Zhang & Tran Trung Luu collected the data; Jicai Zhang performed the analysis and simulations based on the two-level quantum model proposed by Frank Lengers, Daniel Wigger, Doris E. Reiter, & Tilmann Kuhn; Jicai Zhang & Tran Trung Luu performed the data fitting and time-frequency analysis; Ziwen Wang & Jicai Zhang conducted the electronic and phononic related calculations from first principles; All authors contributed to the writing of the manuscript which was first drafted by Jicai Zhang; Hans Jakob Wörner & Tran Trung Luu supervised the project. All authors discussed and interpreted the experimental data.

**Competing interests:** All authors declare that they have no competing interests.

**Data and materials availability:** All data are available in the main text or the Extended Data materials.

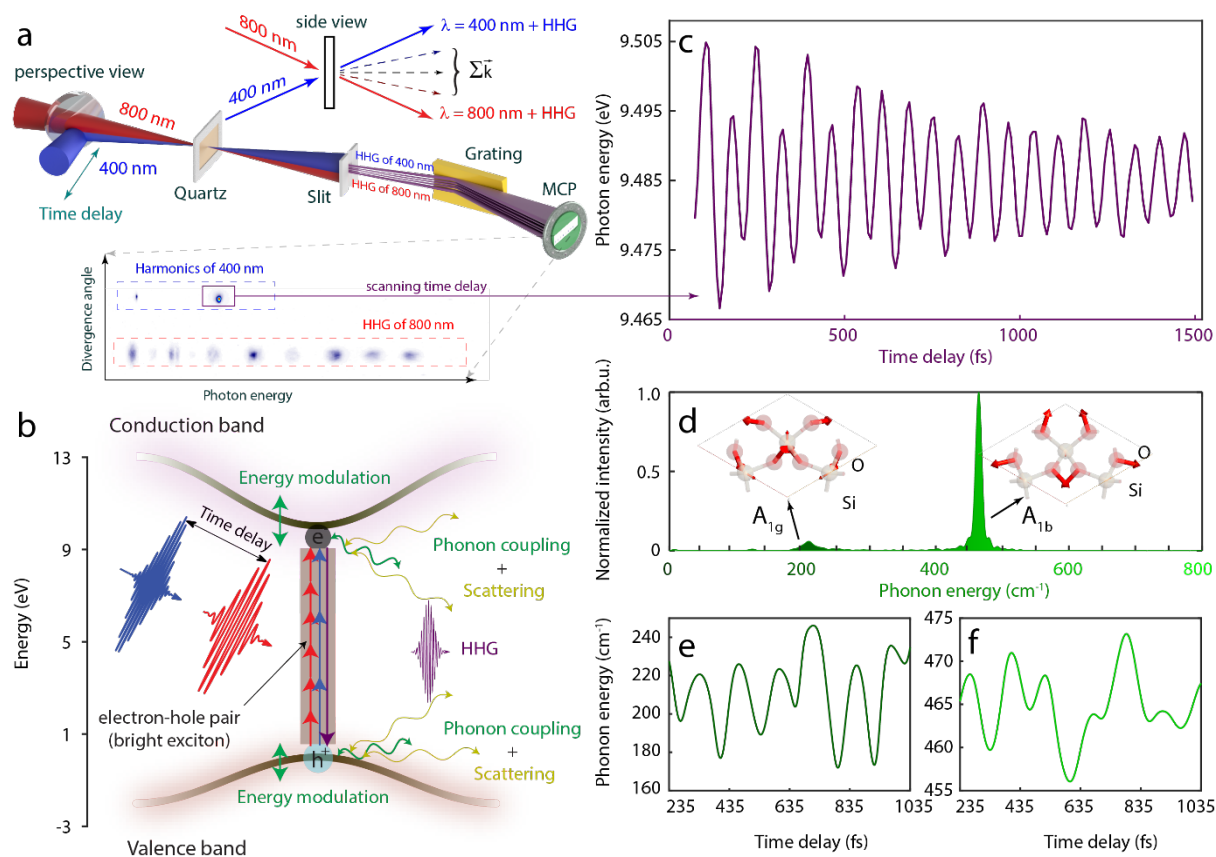

**Fig. 1 | The HHS based principle of measuring lattice vibration dynamics. a,** Schematic of the experimental setup: two ultrashort laser pulses with centre wavelengths at 800 nm (Pump ~30 fs) and 400 nm (Probe ~25 fs) are used to form a non-collinear pump-probe HHG detection geometry. The HHG spectra both from the pump and probe are recorded by an EUV spectrometer placed downstream of the sample. **b**, Optical excitation and *e-ph* interaction scheme in the electron band structure. The red, blue, and violet waves and arrows show the pump excitation, probe excitation, and THG emission respectively. The green and yellow waves represent the phonon creation and scattering. The back-action of the phonons leads to modulations of the electron bands (green arrows). **c**, Detected lattice vibrations from the time-delayed THG spectrum of the probe. **d**, FFT of (**c**), two optical phonons with $A_1$ symmetry located at ~207 cm$^{-1}$ and ~464 cm$^{-1}$ are found and assigned to the $A_{1g}$ and $A_{1b}$ phonon modes with eigenmodes depicted as insets. **e** and **f**, Frequency modulated spectra of $A_{1g}$ (**e**) and $A_{1b}$ (**f**) phonon modes from a time-frequency analysis of **c**.

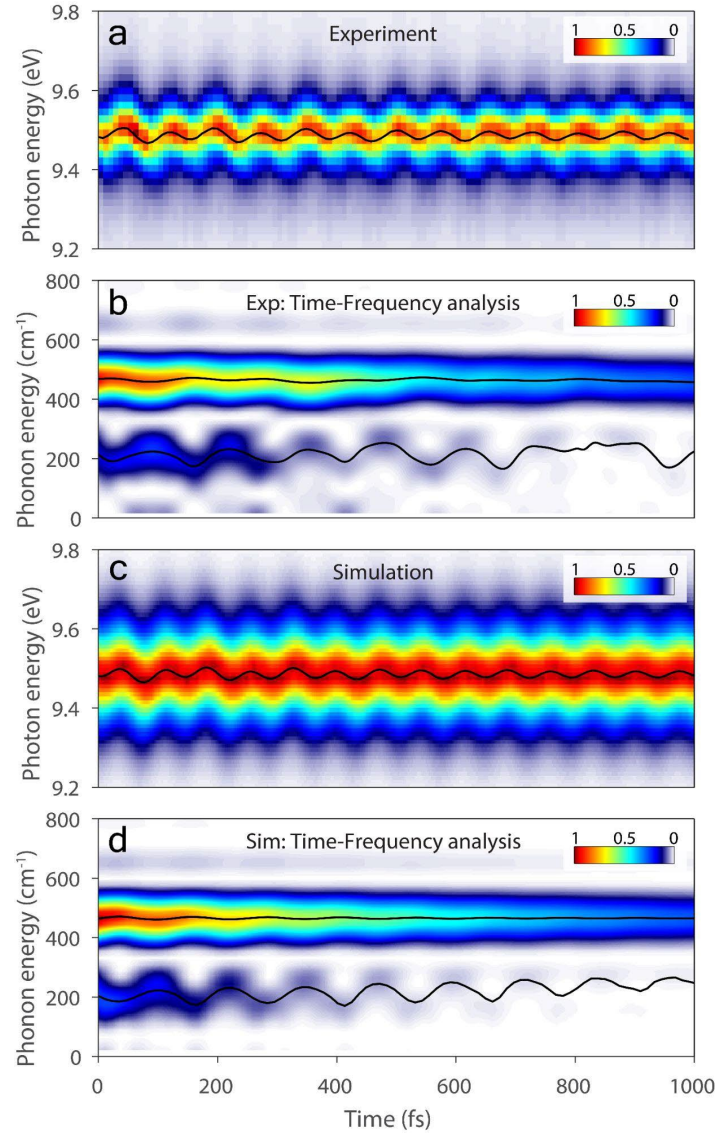

**Fig. 2 | Experimental and simulated time delayed THG trace of lattice dynamics from HHS. a,** Experimentally measured raw data of the integrated time-delayed THG trace of the probe polarized in Γ-K direction. **b**, Time-frequency analysis of a, the two dominant hot phonons show a periodic frequency modulation in their centre frequencies around 207 and 464 cm$^{-1}$. **c**, Reconstructed THG trace that is based on the two-level quantum-classical model summarized in the main and methods text. **d**, Time-frequency analysis of **c**. All solid black lines in the figure represent the COM of the corresponding trace and the intensities of all traces are normalized to their maximum. The colorbars are linear and normalized to the maximal value of the harmonic signal.

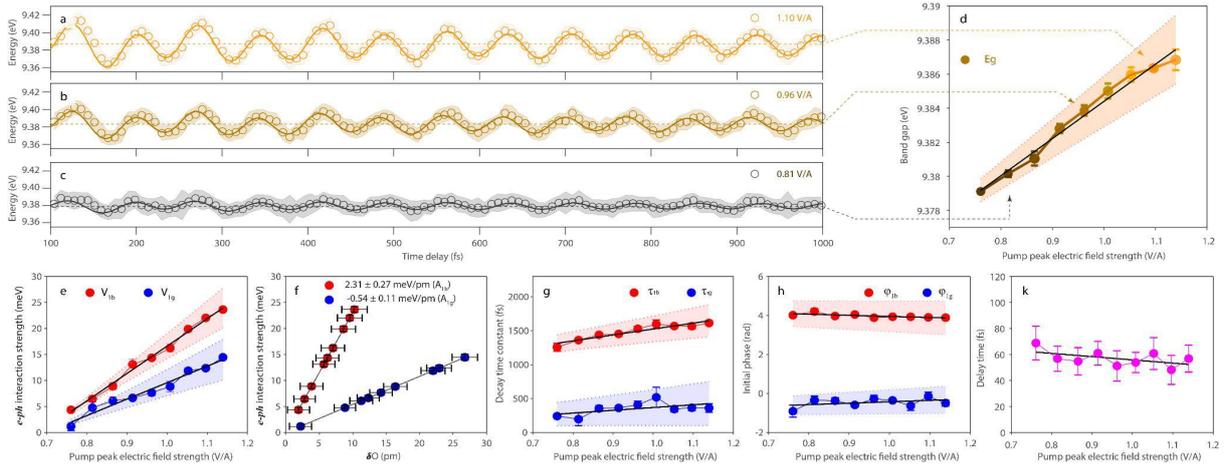

**Fig. 3 | Optical manipulation of electronic and phononic properties of α-quartz crystal. a-c**, Integrated time delayed THG spectra of the probe through the intensity-scaling measurement when both pump and probe pulses are polarized along the Γ−K direction. The estimated pump peak electric field ranging from 0.76 to 1.14 V/Å, and for clarity only three spectra are shown. The dots and solid line represent the experimental measurement and their fits ($R^2$ is around 0.85 to 0.95 from lower to higher laser intensity). **d**, The extracted pump-intensity dependent band gap variation based on the two-level model at vanishing delay time. **e**, same as d but showing the extracted *e-ph* interaction strength variation of the $A_{1b}$ (red dots) and $A_{1g}$ (blue dots) phonon mode. **f** the calculated oxygen atom displacement in the optical phonon modes based on the optical deformation potential calculated by DFPT and the measured *e-ph* interaction strength. **g**, Decay time variation of the two phonon modes. **h** and **k**, Extracted initial phase of the phonon excitation, and corresponding relative delay excitation times. All solid lines from **d** to **k** indicate linear fittings of the extracted data and the error bars denote fitting uncertainties, whereas the shaded areas represent systematic fluence and statistical uncertainties.

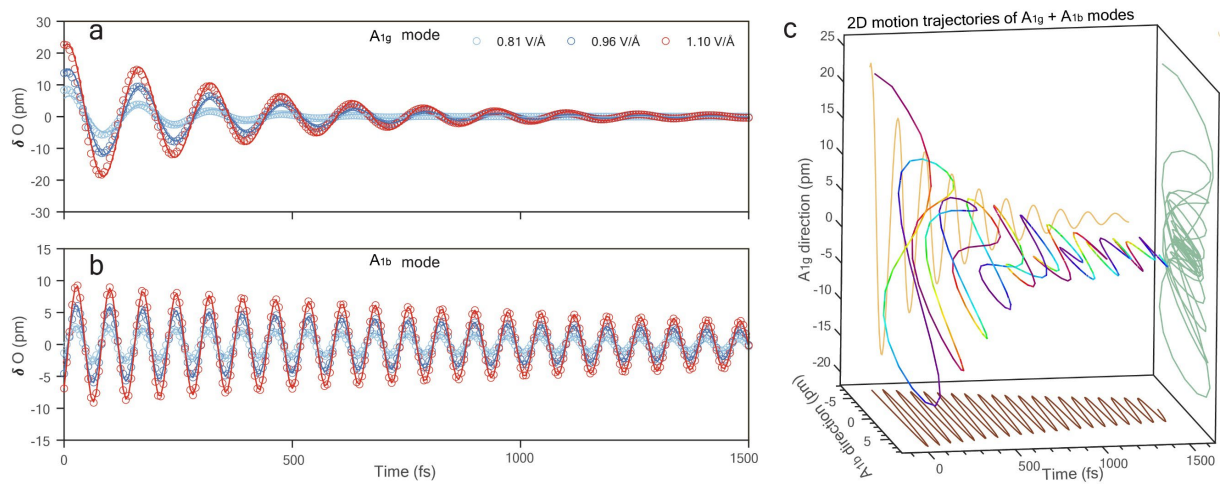

**Fig. 4 | Reconstructed O atom displacement dynamics of the two optical modes $A_{1g}$ and $A_{1b}$ by combining the calculated ODP from DFPT. a**, **b,** The real-time dynamics of the oxygen atomic displacements of the α-quartz crystal by the optical phonons $A_{1g}$, $A_{1b}$. The colored dots represent the different pump strengths of 0.81, 0.96 and 1.10 V/Å. The solid lines indicate the fits of the oscillations. **c**, The combined ($A_{1g} + A_{1b}$) oxygen oscillations exhibit 2D motion trajectories in real space (Cartesian coordinates) when a pump intensity of 1.10 V/Å is selected. The brown, yellow, and green curves represent the projected trajectories of the time-dependent 2D motion in real space.

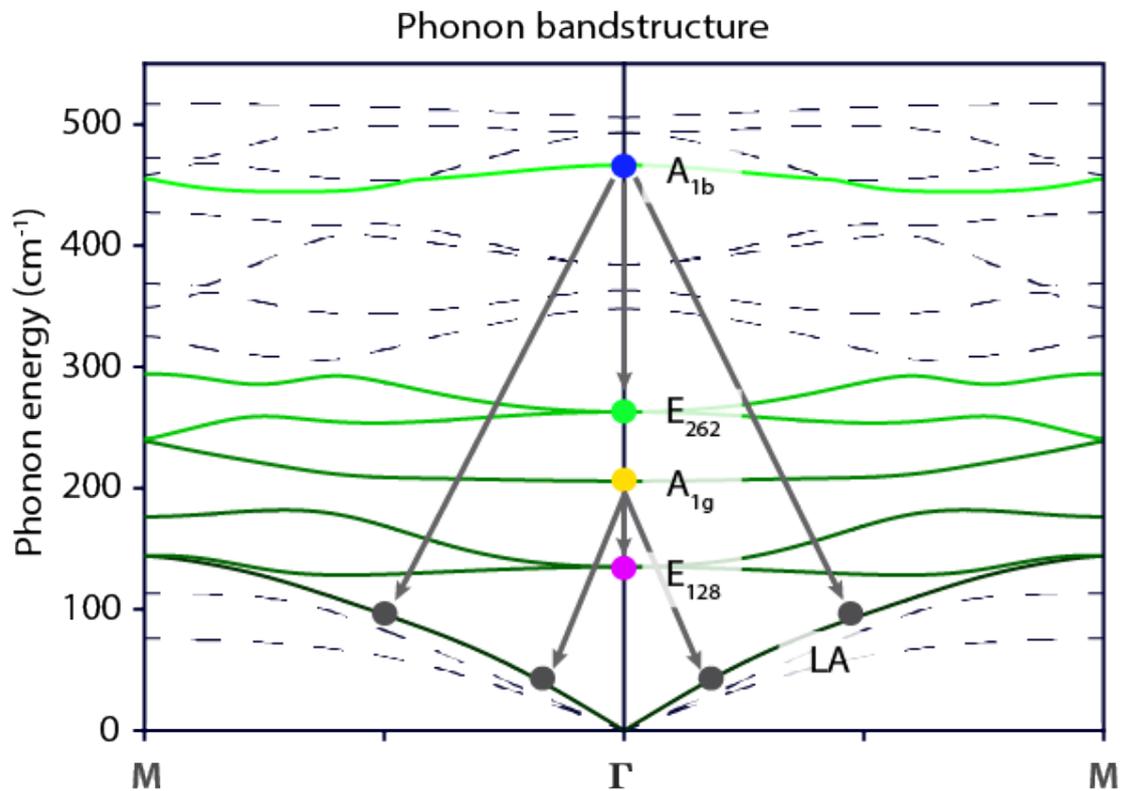

**Fig. 5 | Phonon dispersion and possible anharmonic decays for $A_{1g}$ and $A_{1b}$ phonons by a 4-phonon scattering pathways.** Phonon band structure of α-quartz calculated by DFPT, where the coloured lines represent the participating phonon branches, and the coloured dots represent the optical phonons, and the black dots represent specific ac phonons. The arrows indicate the possible phonon-phonon scattering paths of the the $A_{1g}$ and $A_{1b}$ modes.

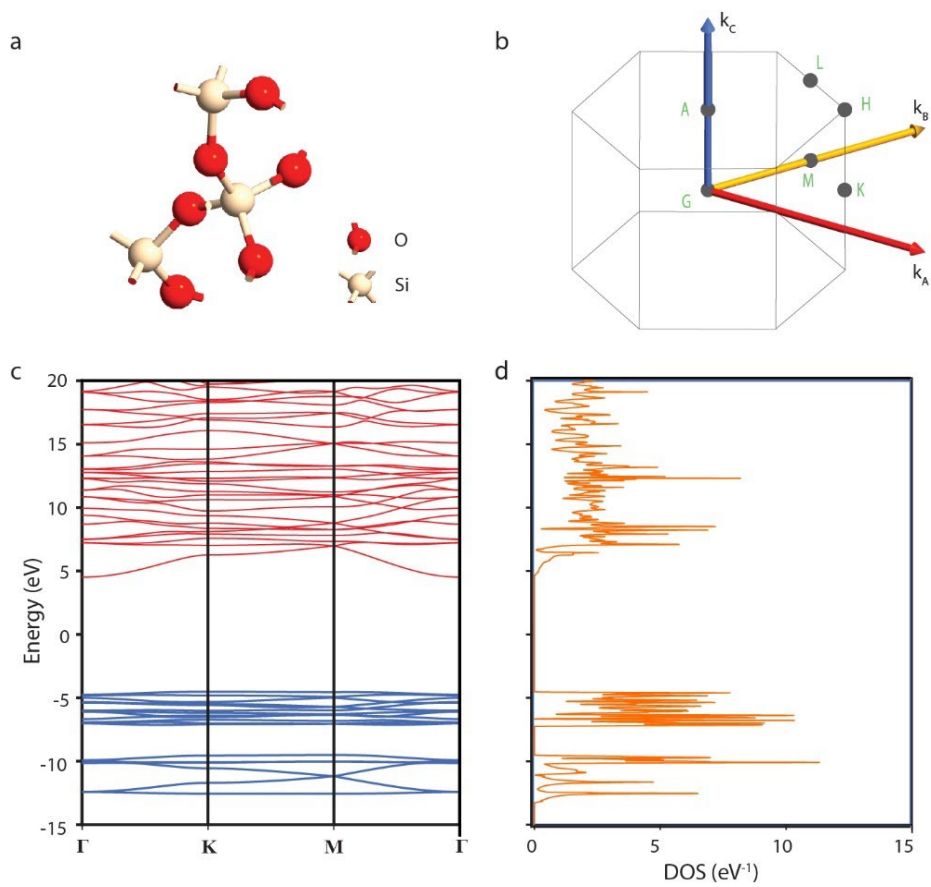

**Extended Data Fig. 1 | Electronic Band structure of quartz crystal. a**, Unit cell of bulk α-quartz crystal which consists of nine atoms (three Si and six O atoms) in total and **b**, the corresponding Brillouin zone. **c** and **d**, Calculated electronic band structure and density of states from DFT.

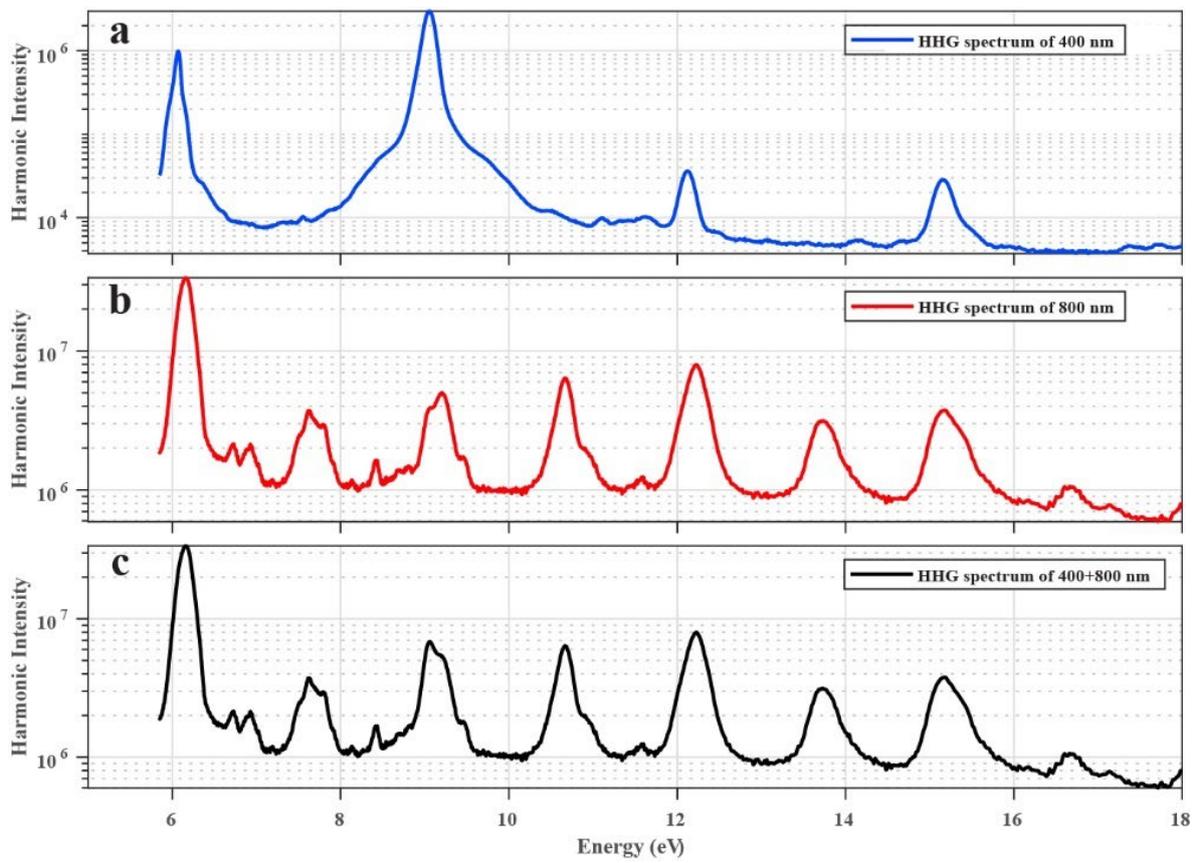

**Extended Data Fig. 2 | Energy-calibrated, spatially resolved HHG spectra. a**, 400 nm only. **b**, 800 nm only. **c**, Combination of them after vertical integration on the MCP image.

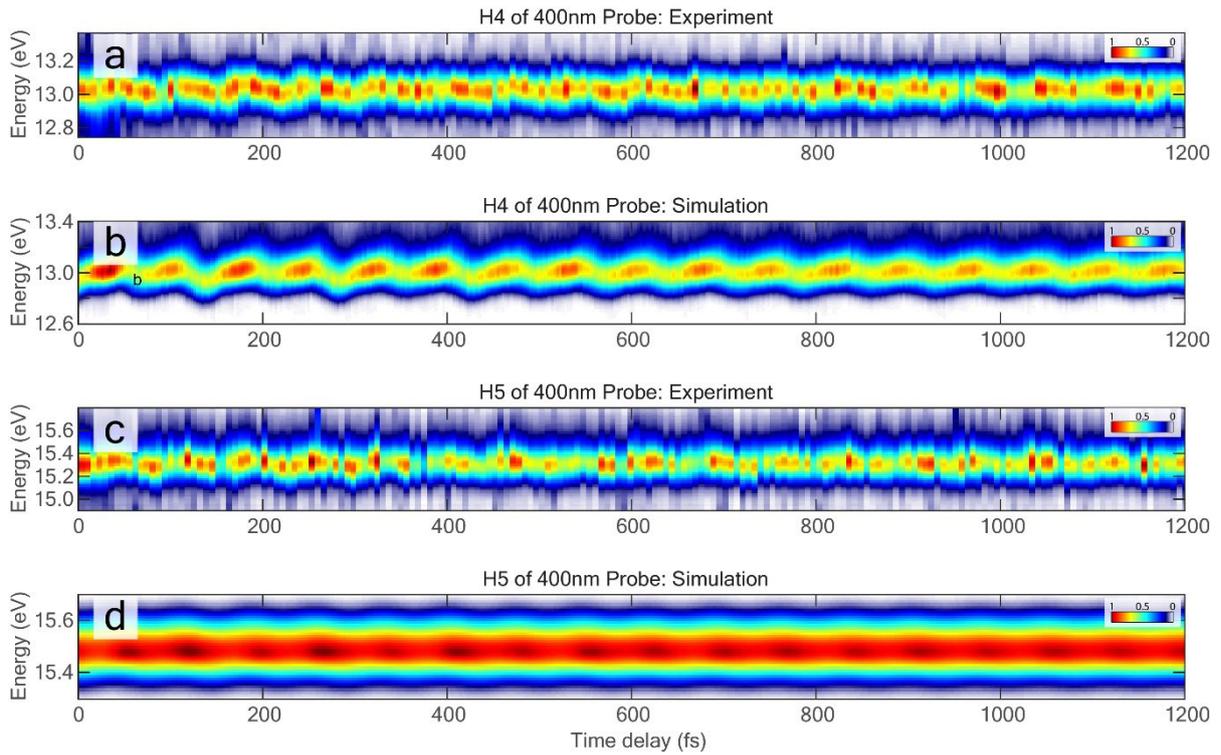

**Extended Data Fig. 3 | Measured and Simulated Time-resolved HHG spectra of 400nm pulses. a** and **c**, Time-resolved spectrum trace of 4$^{th}$ and 5$^{th}$ harmonic of 400 nm probe. **b**, and **d**, The corresponding simulations from the quantum model. The color bar is linear and normalized to the maximal value of the harmonic signal.

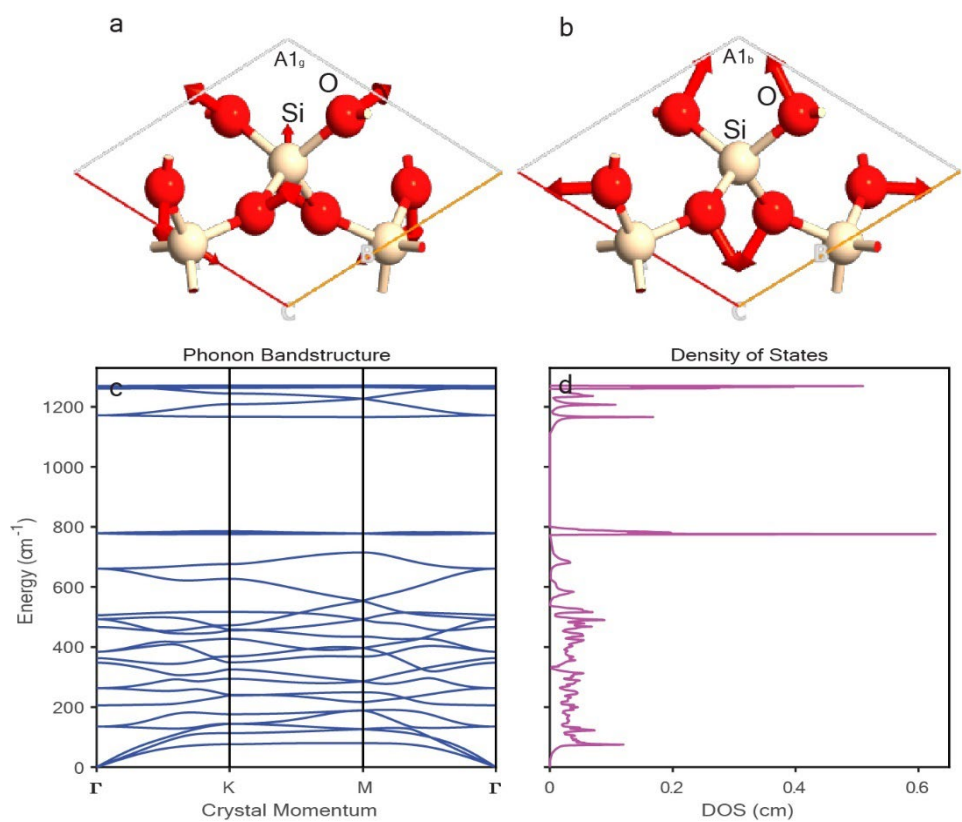

**Extended Data Fig. 4 | Phonon mode and phonon dispersion curve of quartz crystal. a** and **b**, Vibrational modes of $A_{1g}$ and $A_{1b}$ phonon modes of α-quartz in the *z*-direction of the lattice, the arrows and their lengths indicate the directions of atomic motion and oscillation amplitudes. **c** and **d**, Phonon dispersion and phonon density of states that were calculated from DFPT.

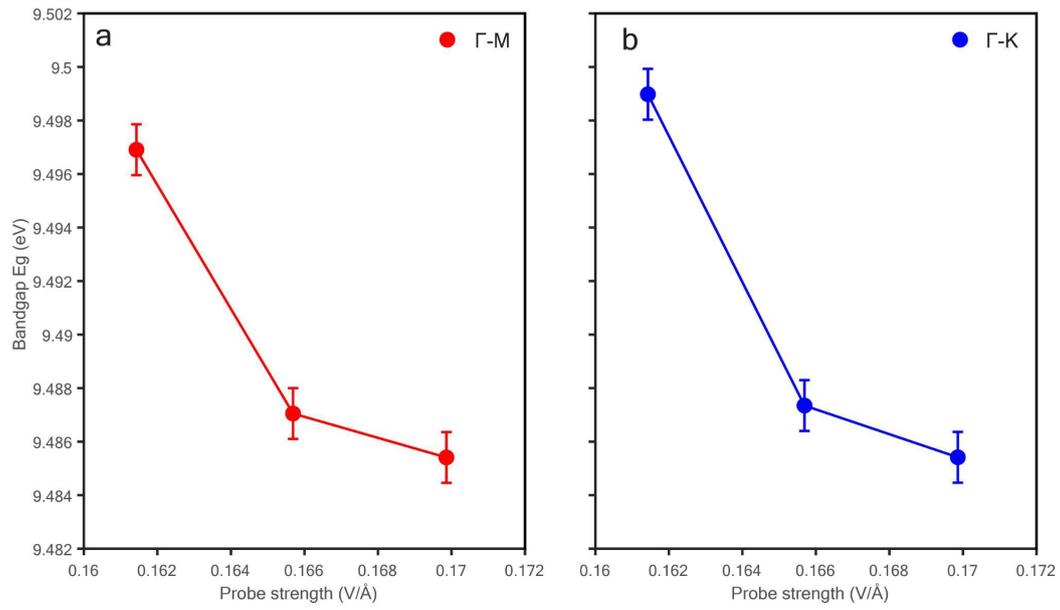

**Extended Data Fig. 5 | Band gap variation at different probe pulse field strengths. a** and **b**, Denote the values extracted from experimental spectra that were measured in Γ-M and Γ-K direction when the pump intensity was fixed aroung 1.1 V/Å, respectively.

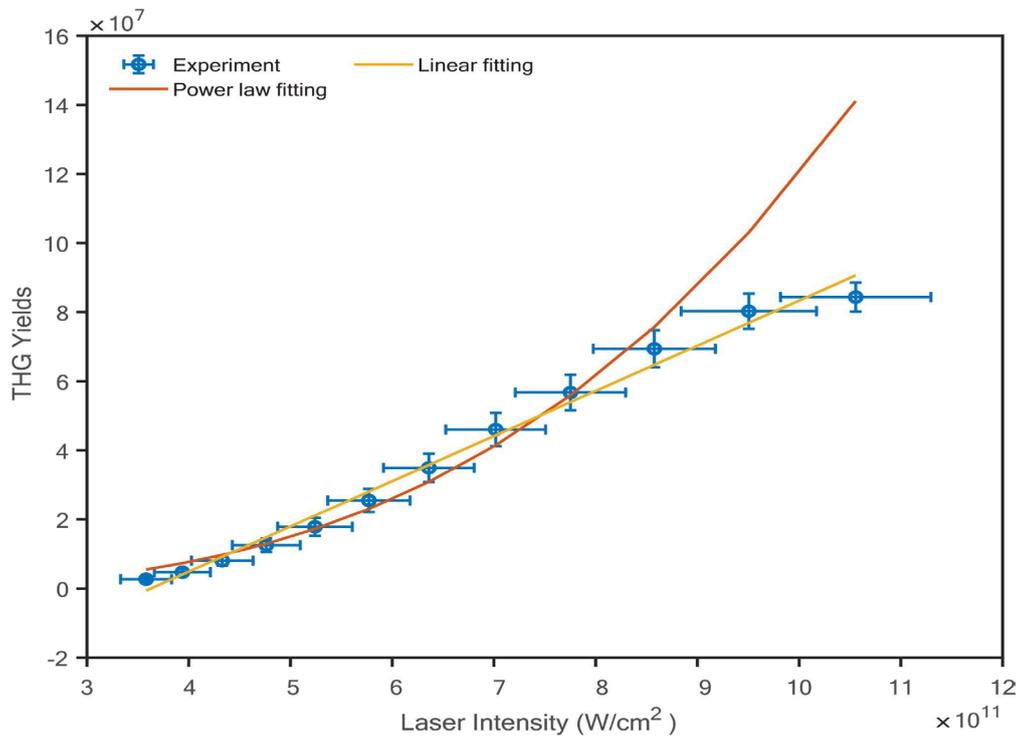

**Extended Data Fig. 6 | Intensity scalling of THG of 400nm probe.** The blue dot line represents the measured THG yields at varying laser intensities, while the red solid line corresponds to the fitting results obtained using a perturbative power law model $I^n$ (I is the pulse intenstity and n is the harmonic order, here n = 3). The yellow line is a linear fit to the experimental data.

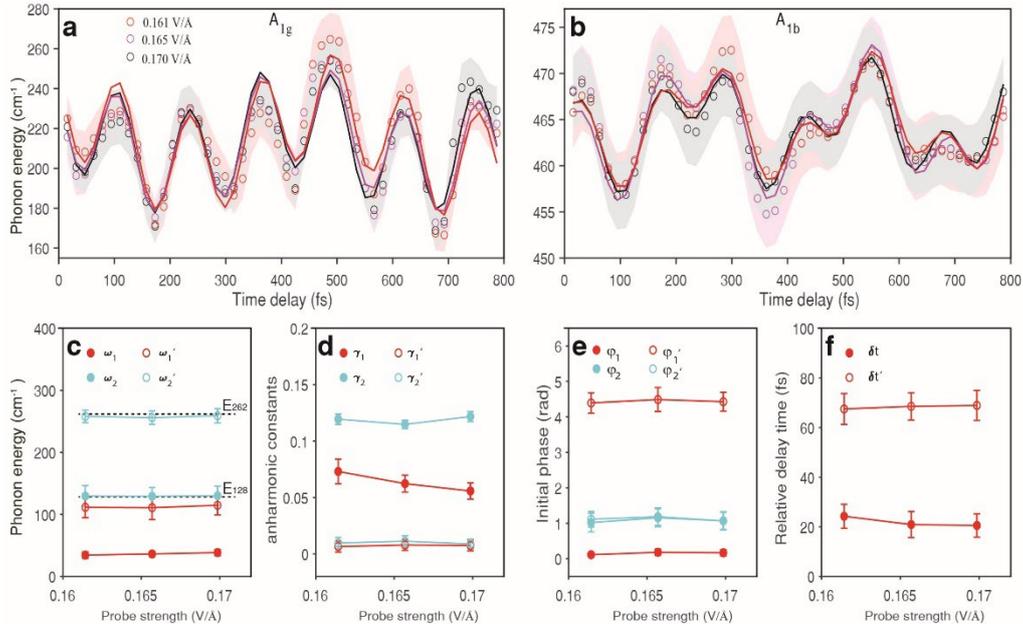

**Extended Data Fig.7 | Quantifying *ph-ph* couplings from the classical coupled oscillator model. a** and **b**, Modulated spectrum of $A_{1g}$ and $A_{1b}$ modes with probe intensity-scaling measurement when both pump and probe pulses are polarized in Γ−M direction. The legends show the estimated probe peak intensity ranging from 0.161 to 0.170 V/Å, and fixed pump field strength ~1.0 V/Å. The solid line indicates the linear fitting according to the coupled classical oscillator interaction model. **c**, Extracted modulation frequency variations for both modes, the dashed lines represent the experimentally observed $E_{128}$ and $E_{262}$ phonon modes[15]. **d**, Anharmonic dimensionless *ph-ph* interaction constants. **e** and **f**, Initial phases of the modulation frequencies and corresponding relative delay time of connected phonon modes. The error bars in c to f include fitting, statistical and systematic uncertainties.

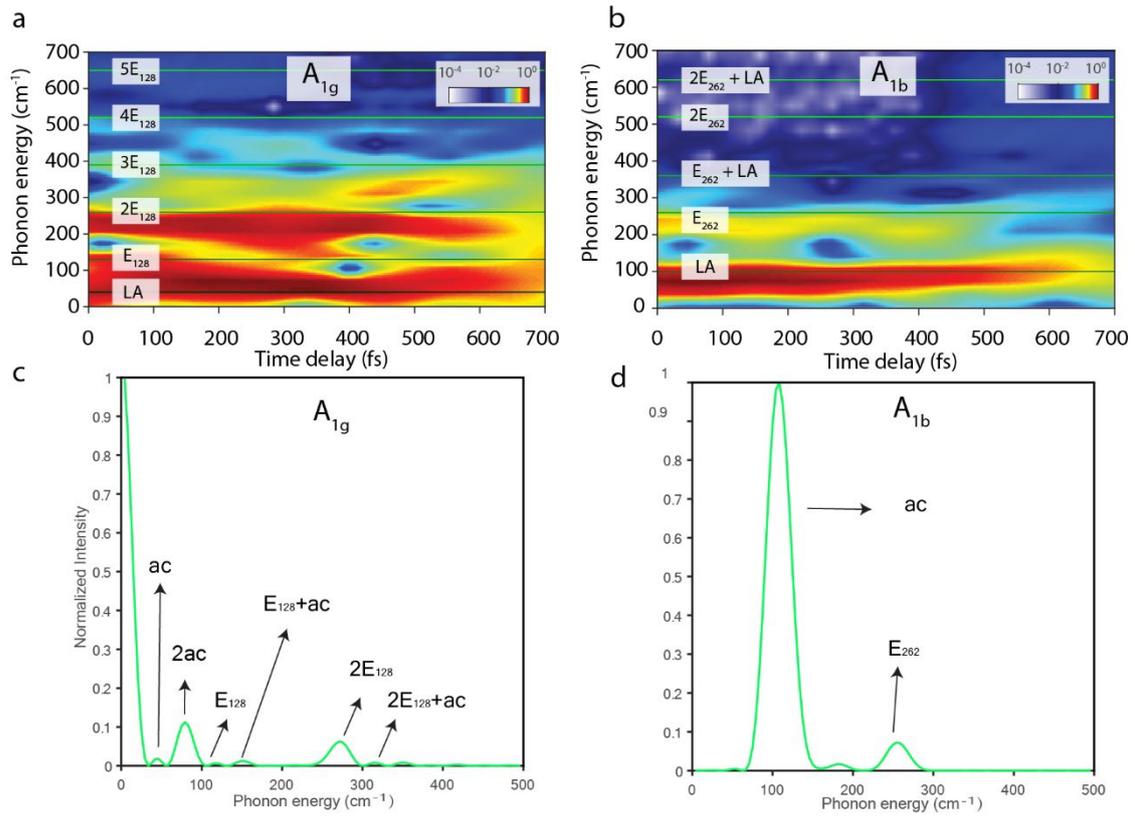

**Extended Data Fig. 8 | Time-frequency and FFT analysis of the modulated spectra. a**. Time-frequency analysis of the modulated spectra from Extended Data Fig. 8**a** and 8**b** in the Γ−M direction. The solid green lines indicate the spacing frequencies of the specific phonon modes that contribute to the four-phonon scattering process. The intensity of all traces is normalized to their maximum. **c** and **d**, direct FFT of the modulated spetra in Fig. 8**a** and 7**b**, respectively. The arrows indicate the main peaks of the spectra.

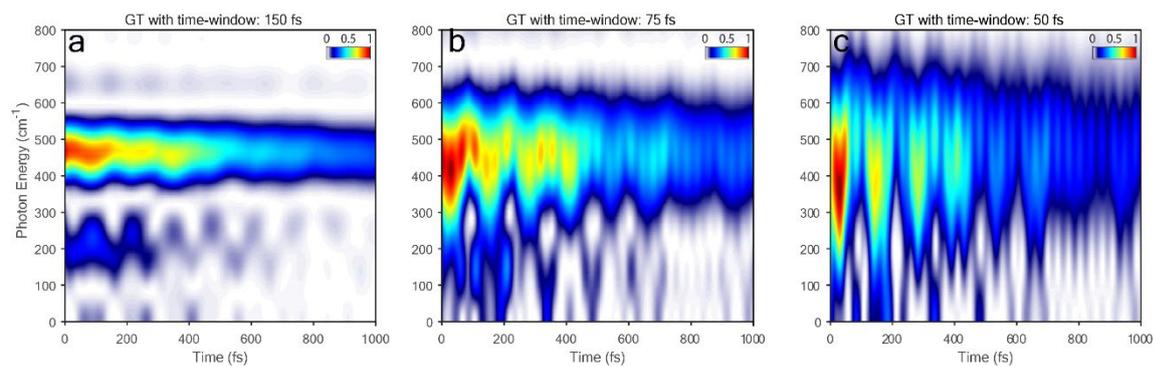

**Extended Data Fig. 9 | Time-Frequency analysis (GT) under different time window widths.  a, b**, and **c.** The GT Spectrum variation under different time window widths from 150 f, 100 fs, and 50 fs, respectively.

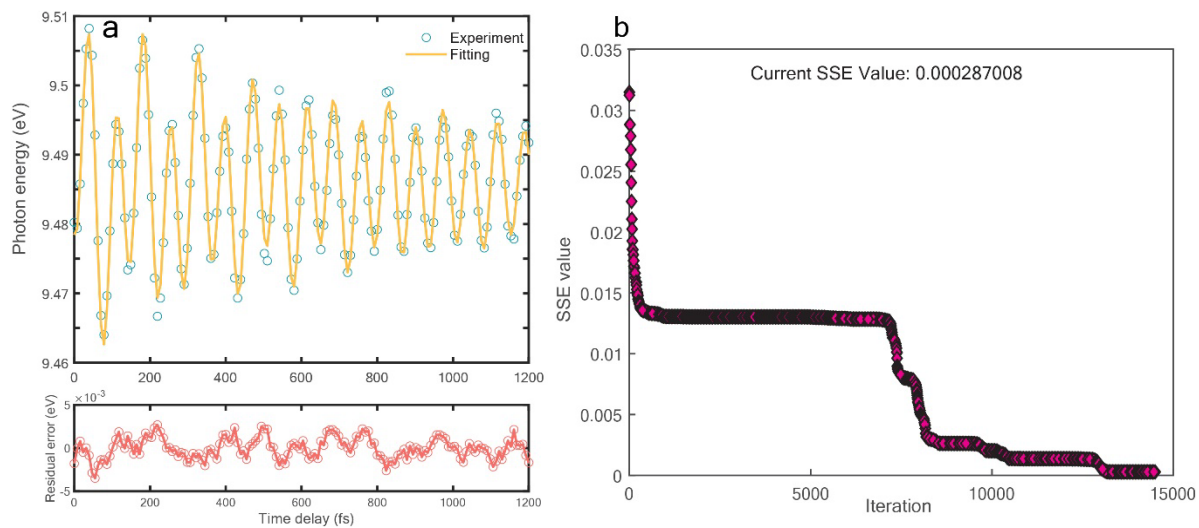

**Extended Data Fig. 10 | Phonon parameters fitting and residual errors. a.** Measured (dot) and fitted (solid line) COM spectrum of THG (upper panel), and the lower panel represents the residual error between the fitted and the measured data. **b.** Iterated SSE value variations after setting four pre-values off by 50 %.

| Symmetry | Phonon Frequency (cm$^{-1}$) | | |
|---|---|---|---|
| A1 | 207 | 205.9 | 205.7 |
| | 356 | 355.6 | 346.8 |
| | 464 | 464.1 | 466.2 |
| | 1085 | 1084.5 | 1171 |
| E(LO+TO) | 128 | 127.9 | 134.7 |
| E(LO+TO) | 265 | 264.6 | 262.3 |
| E(LO+TO) | 697 | 696 | 658.3 |
| E(LO+TO) | 1162 | 1161.5 | 1152 |
| References | [65] | [66] | This work |

**Extended Data Tab. 1 | Comparison of the phonon frequencies.** (A and E symmetry) at the Γ point based on DFPT calculations in this work and the previous experimental Raman spectroscopy measurements.

| Phonon parameters | $V_{1b}$ (meV) | $\omega_{1b}$ (cm$^{-1}$) | $\tau_{1b}$ (fs) | $\varphi_{1b}$ (rad) | $E_g$ (eV) |
|---|---|---|---|---|---|
| Fitted value | 17.2 | 464.2 | 1600.2 | 3.7 | 9.48 |
| Phonon parameters | $V_{1g}$ (meV) | $\omega_{1g}$ (cm$^{-1}$) | $\tau_{1g}$ (fs) | $\varphi_{1g}$ (rad) | |
| Fitted value | 10.3 | 207.1 | 498.2 | 0.22 | |

**Extended Data Tab. 2 | Fitted Phonon parameters of A$_{1b}$ and A$_{1g}$ modes.** The fitted phonon parameters of extended data Fig. 10 are based on equation (3) in the main text.

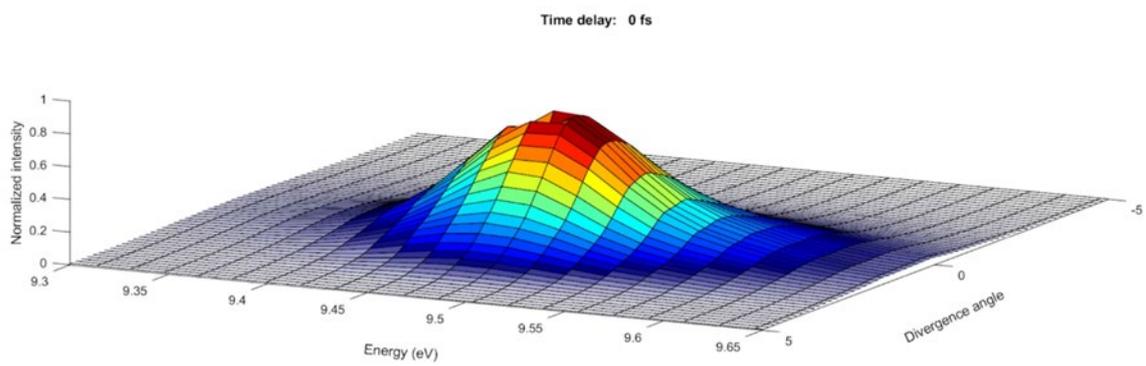
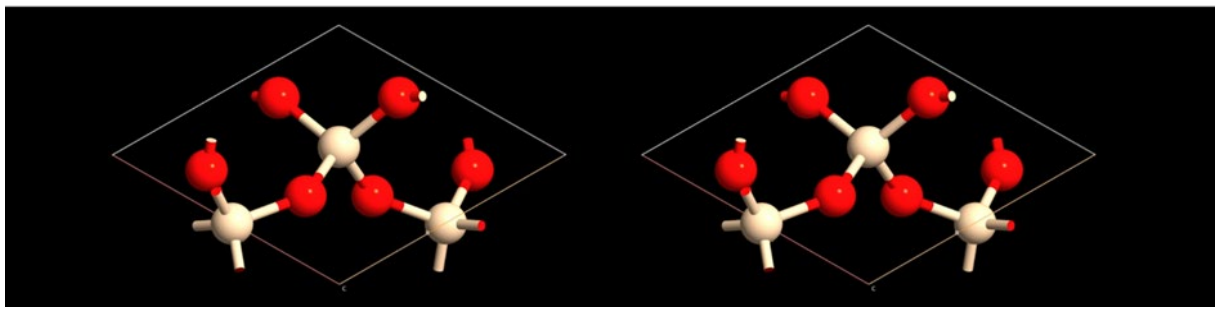

**Extended Data Movie. 1 | Visualizing the lattice vibrations with HHS technique. Movie (screenshot).**